\documentclass[twocolumn,apj,floatfix,showkeys]{emulateapj}
\usepackage{latexsym}
\usepackage{epsfig}
\usepackage{amssymb}
\usepackage{amsmath}
\usepackage[latin1]{inputenc}
\usepackage{amsmath}
\usepackage{pstricks}
\usepackage{graphicx}
\usepackage{hyperref}
\begin{document}


\title{Environment Dependence of Dark Matter Halos in Symmetron Modified Gravity}
\author{Hans A. Winther$^{1,~\hyperref[mail]{\omega}}$}
\author{David F. Mota$^{1,~\hyperref[mail]{\mu}}$}
\author{Baojiu Li$^{2,3,4,5~\hyperref[mail]{\lambda}}$}
\affiliation{$^{1}$Institute of Theoretical Astrophysics,
University of Oslo, 0315 Oslo, Norway} 
\affiliation{$^{2}$ICC, Department of Physics, University of Durham, South Road, Durham DH1 3LE, UK}
\affiliation{$^{3}$DAMTP,
Centre for Mathematical Sciences, University of Cambridge,
Wilberforce Road, Cambridge CB3 0WA, UK}
\affiliation{$^{4}$Kavli
Institute for Cosmology Cambridge, Madingley Road, Cambridge CB3
0HA, UK}
\affiliation{$^{5}$Institute of Astronomy, Madingley Road, Cambridge CB3 0HA, UK}

\footnotetext[]{\label{mail}
\\$^{\omega}$	Email address: \href{mailto:h.a.winther@astro.uio.no}{\nolinkurl{h.a.winther@astro.uio.no}}
\\$^{\mu}$		Email address: \href{mailto:d.f.mota@astro.uio.no}{\nolinkurl{d.f.mota@astro.uio.no}}
\\$^{\lambda}$	Email address: \href{mailto:B.Li@damtp.cam.ac.uk}{\nolinkurl{b.li@damtp.cam.ac.uk}}}


\begin{abstract}
We investigate the environment dependence of dark matter halos in the symmetron modified gravity scenario. The symmetron is one of three known mechanisms for screening a fifth-force and thereby recovering General Relativity (GR) in dense environments. The effectiveness of the screening depends on both the mass of the object and the environment it lies in. Using high-resolution N-body simulations we find a significant difference, which depends on the halo's mass and environment, between the lensing and dynamical masses of dark matter halos similar to the $f(R)$ modified gravity. The symmetron can however yield stronger signatures due to a freedom in the strength of coupling to matter.
\end{abstract}


\maketitle


\section{Introduction}
Many theories of high energy physics, like string theory and supergravity, predict
light gravitationally coupled scalar fields (see e.g.
\cite{supersymmetry_review,linde} and references therein). These
scalars may play the role of dark energy (quintessence). If these
scalar fields have non-minimal coupling to matter fields, then
they could mediate extra forces and are potentially detectable
in local experiments and from observations on cosmological scales.

Several laboratory and solar system
experiments over the last decades have tried to detect a sign of such fundamental
coupled scalar fields \citep{eotwash,hoskins,decca,cassini}, but
the results so far show no signature of them. Naively, the results
of these experiments have ruled out any such scalar fields, which also 
can have an effect on the large-scale structure of the Universe, unless 
there is some mechanism which suppresses the scalar fifth-force on small scales where the experiments are performed. One should keep in mind that GR is only well tested on length scales ranging from millimeters to the size of the solar system. Comparing this to the size of the horizon, this leaves a wide range of scales where there could be deviations from GR.

To this day we know three such types of theoretical mechanisms (see
\cite{khoury_screening_mechanisms} for a review) that can explain
why light scalars, if they exist, may not be visible in experiments performed near the Earth. One such class, the
chameleon mechanism
\citep{chameleon_cosmology,cosmological_chameleon,alpha3,alpha1,alpha2,brax_generalchameleon,mota_shaw}, operates when
the scalars are coupled to matter in such a way that their
effective mass depends on the local matter density. In regions
where the local mass density is low, the scalars would be light
and deviations from GR would be observed. But near
the Earth, where experiments are performed, the local mass density
is high and the scalar field would acquire a heavy mass making the
interactions short range and therefore unobservable. This mechanism is 
the reason why $f(R)$ modified gravity can lead to viable cosmologies and still evade local gravity constraints.

The second mechanism, the Vainshtein mechanism
\citep{Vainshtein,dvali_gabadadze_vains,schwartz_massivegravity},
operates when the scalar has derivative self-couplings which
become important near matter sources such as the Earth. The strong
coupling near sources essentially cranks up the kinetic terms,
which translates into a weakened matter coupling. Thus the scalar
screens itself and becomes invisible to experiments. This
mechanism is central to the phenomenological viability of
braneworld modifications of gravity and galileon scalar theories
\citep{dgp,cascading_dgp,galileon_modifiedgravity,multifield_galileons,gal1,gr_aux_dim,massivegravity,davis_gal}.

The last mechanism, the one explored in this paper, is the
symmetron mechanism
\citep{khoury_symmetron,symmetron_cosmology,olive_pospelov,2011arXiv1108.3082B,chameleon_signatures}. In
this mechanism, the vacuum expectation value (VEV) of the scalar
depends on the local matter density, becoming large in regions of
low mass density, and small in regions of high mass density. The scalar couples with
gravitational strength in regions of low density, but is decoupled
and screened in regions of high density. This is achieved through
the interplay of a symmetry breaking potential and a universal
quadratic coupling to matter. A similar screening mechanism applies for the environmentally dependent dilaton model \cite{brax_dilaton}.

In vacuum, the scalar acquires a nonzero VEV which spontaneously breaks
the $\mathbb{Z}_2$ symmetry $\phi\to-\phi$. In the regions of
sufficiently high matter density, the field is confined near $\phi
= 0$, and the symmetry is restored. The fifth force arising from
the matter coupling is proportional to $\phi$, making the effects
of the scalar small in high density regions. Because of this effect, dark matter halos in 
high-density regions will produce different scalar fifth forces compared to those in low-density regions.

This effect has been studied for the case of $f(R)$ gravity (Chameleon mechanism) in \cite{2011PhRvL.107g1303Z,2010PhRvD..81j3002S} and the Dvali-Gabadadze-Porratti (DGP) model (Vainshtein mechanism) in \cite{2010PhRvD..81j3002S}. It was found that in the DGP model the screening of halos is almost independent of environment while in $f(R)$ gravity there can be a significant environmental dependence. Another signature that has been found recently in $f(R)$ and symmetron models is that the luminosity of galaxies \citep{davis_galaxies} might also depend on the environment.

Recent work on the symmetron has focused on background cosmology, linear \cite{2011arXiv1108.3082B} and nonlinear structure formation \cite{2011arXiv1108.3081D}, and also made halo scale predictions \cite{2011arXiv1110.2177C}. In this paper we use the high-resolution N-body simulations of \cite{2011arXiv1108.3081D} to study the environment dependence of dark matter halos in the symmetron modified gravity scenario (see for instance \cite{baojiu_fofr,li1,li2,baojiu_extended_quintessence,schmidt_fofr1,schmidt_fofr2,schmidt_dgp,2009PhRvD..80h3522H,dilaton} for N-body simulations within other scenarios of modified gravity).

This paper is organised as follows. In Section~(\ref{sec1}) we recall the main properties of the symmetron model which are relevant for our analysis. Then, in Section~(\ref{sec2}) we introduce the dynamical and lensing masses of a halo and explain how these are obtained from the N-body simulations. In Section~(\ref{env}) we explain how we define the environment of a halo in our analysis. The main results are shown and discussed in Section~(\ref{sec4}), and we also compare our simulation results to semi-anaytical predictions in Section~(\ref{sec5}). Finally we summarize and give our conclusions in Section~(\ref{sec6}).


\section{Symmetron Review}\label{sec1}
The symmetron modified gravity is a scalar field theory specified by the following action
\begin{align}\label{action}
S =& \int dx^4 \sqrt{-g}\left[\frac{R}{2}M_{\rm pl}^2 - \frac{1}{2}(\partial\phi)^2 -
V(\phi)\right]\nonumber\\
&+ S_m(\tilde{g}_{\mu\nu},\psi_i),
\end{align}
where $g$ is the determinant of the metric $g_{\mu\nu}$, $R$ is
the Ricci scalar, $\psi_i$ are the different matter fields and
$M_{\rm pl} \equiv 1/\sqrt{8\pi G}$ where $G$ is the bare
gravitational constant. The matter fields couple universally to the
Jordan frame metric $\tilde{g}_{\mu\nu}$, which is a conformal rescaling
of the Einstein frame metric $g_{\mu\nu}$ given by
\begin{equation}\label{conformal_coupling}
\tilde{g}_{\mu\nu} = A^2(\phi)g_{\mu\nu}.
\end{equation}
The equation of motion for the symmetron field $\phi$ following from the action Eq.~(\ref{action}) reads
\begin{equation}\label{eom_phi}
\square\phi = V_{,\phi} + A_{,\phi}\rho_m \equiv V_{\rm eff,\phi},
\end{equation}
where the potential is chosen to be of the symmetry breaking form
\begin{equation}\label{potential}
V(\phi) = -\frac{1}{2}\mu^2\phi^2 + \frac{1}{4}\lambda\phi^4,
\end{equation}
and the coupling is quadratic in $\phi$ to be compatible with the $\phi\to -\phi$ symmetry
\begin{equation}\label{coupling_function}
A(\phi) = 1 + \frac{1}{2}\left(\frac{\phi}{M}\right)^2.
\end{equation}
The effective potential can then be written as
\begin{align}\label{veff}
V_{\rm eff}(\phi) &= \frac{1}{2}\left(\frac{\rho_m}{M^2}-\mu^2\right)\phi^2 + \frac{1}{4}\lambda\phi^4,
\end{align}
from which the range of the scalar field (i.e., the range of the resulting fifth-force) can be found as
\begin{equation}
\lambda_{\phi} \equiv \frac{1}{\sqrt{V_{\rm eff,\phi\phi}}}.
\end{equation}
The range of the field in vacuum, denoted $\lambda_0$, is given by $\lambda_0 = 1/\sqrt{2}\mu$.

In high-density regions where $\rho_m > \mu^2M^2$ the effective potential has a minimum at $\phi = 0$. The fifth-force, given by 
\begin{equation}
\vec{F}_{\phi} = \vec{\nabla}A(\phi) = \frac{\phi}{M^2}\vec{\nabla}\phi,
\end{equation}
is proportional to $\phi$ and will be suppressed. In vacuum, or in large underdensities, where $\rho_m \ll  \mu^2M^2$ the $\phi \to -\phi$ symmetry is broken and the field settles at one of the two minima $\phi = \phi_0 \equiv \pm\mu/\sqrt{\lambda}$. The fifth-force between two small test masses in such a region will achieve its maximum value compared to gravity,
\begin{equation}\label{maxff}
\frac{F_{\phi}}{F_N} = 2M_{\rm pl}^2 \left(\frac{d\ln A}{d\phi}\right)_{\phi = \phi_0}^2 = 2\beta^2.
\end{equation}
For very large bodies in the sense that
\begin{align}\label{alpha}
\alpha^{-1} \equiv 2\frac{\rho_{\rm SSB}}{\rho_{\rm body}}\left(\frac{\lambda_0}{R_{\rm body}}\right)^2 \ll 1,
\end{align}
the situation is quite different \citep{khoury_symmetron}. Here the symmetry is restored in the interior of the body and the fifth force on a test mass outside becomes suppressed by a factor $\alpha^{-1}$:
\begin{align}
\frac{F_{\phi}}{F_N} = 2\beta^2\frac{1}{\alpha}.
\end{align}
If the body lies in a high density environment where $\phi = \phi_{\rm env} < \phi_0$ the fifth-force will be further suppressed by a factor $\left(\phi_{\rm env}/\phi_0\right)^2$. Thus there are two ways a body can be screened: it can be large enough as to make $\alpha^{-1} \ll 1$ or it can be located in a high density region where $\phi_{\rm env} \ll \phi_0$. The latter in particular leads to an environment dependence of the fifth-force in a dark matter halo which we will investigate in the next section.

Instead of working with $\mu$, $M$ and $\lambda$ we chose to define three more physically intuitive parameters $L$, $\beta$ and $z_{\rm SSB}$ which are the (vacuum) range of the field in Mpc$/h$, the coupling strength to matter and the cosmological redshift where symmetry breaking takes place on the background level respectively. The conversion to the original model parameters are given by
\begin{align}
L = \frac{\lambda_{0}}{\text{Mpc}/h} &= \frac{3000 H_0}{\sqrt{2}\mu},\\
\beta = \frac{\phi_0 M_{\rm pl}}{M^2} &= \frac{\mu M_{\rm pl}}{\sqrt{\lambda}M^2},\\
(1+z_{\rm SSB})^3 &= \frac{\mu^2 M^2}{\rho_{m 0}}.
\end{align}


\section{The Dynamical and Lensing Masses}
\label{sec2}

In any universally coupled scalar-field theory, like the symmetron, we have the choice of describing the dynamics of the model in two mathematically equivalent frames defined by choosing either $g_{\mu\nu}$ or $\tilde{g}_{\mu\nu}$ in Eq.~(\ref{conformal_coupling}) as the space-time metric. In the {\it Einstein-frame}, the one described by Eq.~(\ref{action}), gravity is described by standard GR, but the geodesic equation is modified compared to GR:
\begin{align}\label{geodet}
\ddot{x}^{\mu} + \Gamma^{\mu}_{\alpha\beta}\dot{x}^{\alpha}\dot{x}^{\beta} = -\frac{d\log A(\phi)}{d\phi}\left(\phi^{,\mu} + 2\phi_{,\beta}\dot{x}^{\beta}\dot{x}^{\mu}\right).
\end{align}
In the {\it Jordan-frame} gravity is described by a scalar-tensor-like modified gravity theory, but the matter particles follow the geodesics of the space-time metric $\tilde{g}_{\mu\nu}$
\begin{align}
\ddot{x}^{\mu} + \tilde{\Gamma}^{\mu}_{\alpha\beta}\dot{x}^{\alpha}\dot{x}^{\beta} = 0,
\end{align}
where $\tilde{\Gamma}$ is the Levi-Civita connection of $\tilde{g}_{\mu\nu}$. The predictions of the theory is usually easier to derive in the Einstein-frame and the corresponding quantities can be found in the Jordan-frame by performing the transformation Eq.~(\ref{conformal_coupling}).

Working in the Einstein-frame and the conformal Newtonian gauge, the line-element can be written as
\begin{equation}
ds^2 = a^2(\eta)\left[-d\eta^2(1+2\Phi_N) + (1-2\Phi_N)d\vec{x}^2\right],
\end{equation}
where $\Phi_N$ is the usual Newtonian potential. Transforming to the Jordan-frame using Eq.~(\ref{conformal_coupling}) we find
\begin{equation}
ds^2 = a^2(\eta)\left[-d\eta^2(1+2\Phi) + (1-2\Psi)d\vec{x}^2\right],
\end{equation}
where
\begin{align}
\Phi & \simeq \Phi_N+ \delta A(\phi),\\
\Psi & \simeq \Phi_N -  \delta A(\phi),
\end{align}
with
\begin{align}
 \delta A(\phi) \equiv A(\phi) - 1 = \frac{1}{2}\left(\frac{\phi}{M}\right)^2. 
\end{align}
Note that we have neglected a term\footnote{For the symmetron it was shown in \cite{2011arXiv1108.3081D} that $ \delta A(\phi) \leq  \delta A(\phi_0) \sim \beta^2 10^{-6}$ which for the values of $\beta \lesssim \mathcal{O}(1)$ we are interested in is much less that one.} $2\Phi_N  \delta A(\phi) \ll \Phi_N$ in the equations above. In the solar-system deviations from GR are often phrased in terms of the so-called PPN parameter $\gamma$. In the case of the symmetron we have
\begin{align}
\gamma = \frac{\Psi}{\Phi} = \frac{\Phi_N -  \delta A(\phi)}{\Phi_N +  \delta A\phi)} = 1 - \frac{2 \delta A(\phi)}{\Phi_N + \delta A(\phi)}.
\end{align}
The solar system constraints for the symmetron were derived in \cite{khoury_symmetron} and gives a constraint on the range of the field and the symmetry breaking redshift: $L(1+z_{\rm SSB})^3 \lesssim 2.3$ \citep{2011arXiv1108.3081D}.

Typically observations of e.g. clusters probes forces (gradients of the potentials) instead of the potentials themselves and different observables are related to different combinations of the potentials. The fifth-force potential is given by the difference in the above two potentials
\begin{align}
\Phi_- = \frac{\Phi - \Psi}{2} =  \delta A(\phi).
\end{align}
Lensing on the other hand is affected by the lensing potential
\begin{align}
\Phi_+ = \frac{\Phi + \Psi}{2} = \Phi_N,
\end{align}
which satisfies the Poisson equation
\begin{equation}
\nabla^2\Phi_+ = 4\pi G a^2 \delta\rho_m.
\end{equation}
This is the same equation as in GR since the action of the electromagnetic field is conformally invariant and thus photons do not feel the scalar fifth-force. In general, there will also be a contribution from the clustering of the scalar-field $4\pi G a^2 \delta V(\phi)$, but in our case this term is negligible as the difference in the clustered and unclustered energy density of the scalar field is always much less than the energy density of matter in a halo\footnote{The potential energy of the scalar field satisfies $|\delta V(\phi)| \leq |V(\phi_0) -V(0)| = \mu^4/4\lambda \sim \rho_{\rm SSB}\beta^2\left(M/M_{\rm pl}\right)^2 \lesssim \overline{\rho}_{m0}\beta^2(1+z_{\rm SSB})^310^{-6}$. For $\beta,z_{\rm SSB} \lesssim \mathcal{O}(1)$ this term is negligible compared to the energy density of matter in a halo.}. We define the lensing mass as
\begin{equation}
M_L = \frac{1}{4\pi G a^2}\int \nabla^2\Phi_+ dV.
\end{equation}
which is the actual mass of the halo. It is determined form the N-body simulations by counting the number of particles within a given radius. For spherical symmetry we can use Stokes theorem, $\int \nabla^2\Phi_+ dV = \int \nabla\Phi_+\cdot d\vec{S} = 4\pi r^2 \frac{d\Phi_+}{dr}$, which gives
\begin{equation}
M_L(r) \propto r^2\frac{d\Phi_+}{dr}.
\end{equation}
The dynamical mass $M_D(r)$ of a halo is defined as the mass contained within a radius $r$ as inferred from the gravitational potential $\Phi$, i.e.
\begin{equation}
M_D(r) = \frac{1}{4\pi G a^2}\int \nabla^2\Phi dV,
\end{equation}
where the integration is over the volume of the body out to radius $r$. For spherical symmetry we can again use Stokes theorem on the r.h.s to find
\begin{equation}\label{dyn_mass}
M_D(r) \propto \int r^2 \frac{d\Phi(r)}{dr} = r^2\left(\frac{d\Phi_N}{dr} + \frac{\phi}{M^2}\frac{d\phi}{dr}\right).
\end{equation}
The terms in the brackets are recognized as the sum of the gravitational force and the fifth-force. In our N-body simulations we measure $M_D$ of a halo by first using a halo-finder to locate the particles which make up the halo and bin them according to radius. Then, we calculate the average total force in each radial bin by summing over all the particles in the bin. Note that the force obtained in this way can have a contribution from the particles outside the halo. For spherical symmetry, this contribution largely cancels out and we are left with the total force produced by the halo itself. Observationally, $M_D$ can be determined from measurements of e.g. the velocity dispersion of galaxies in halos \citep{2010PhRvD..81j3002S}.
\\\\
In GR the lensing mass is the same as the dynamical mass, but they can be significantly different in modified gravity. We follow \cite{2011PhRvL.107g1303Z} and define the relative difference
\begin{equation}
\Delta_M(r) = \frac{M_D}{M_L}-1 = \frac{d\Phi_-/dr}{d\Phi_+/dr}.
\end{equation}
This allows us to quantity the difference between the two masses in the simulations. In GR we have $\Delta_M \equiv 0$ while in the symmetron model $\Delta_M$ will vary depending on the mass of the halo and the environment it lies in. The theoretical maximum is achieved for small objects in a low-density environment where the screening is negligible and reads (see Eq.~\ref{maxff})
\begin{equation}
\Delta_M^{\rm Max}(r) = 2\beta^2.
\end{equation}
In the next section we will apply these to the N-body simulation results.

\section{Defining the environment}\label{env}

The environment a halo lies in can have large effects on the fifth-force that operates by the halo since the fifth-force is directly proportional to the spatial gradient of the square of the local field value (see Eq.~(\ref{dyn_mass})). This value is small in high-density regions and this will provide the halo with an additional screening to the self-screening due to its size or mass.

As a result, when looking for an environmental dependence it is crucial to choose a definition of the environment that does not correlate heavily with the halo mass. The quantity one chooses should also allow for an easy determination both in our simulations and in observations. Such a quantity was found in \cite{2011arXiv1103.0547H} and used in the same analysis as we have done, but for the case of $f(R)$ gravity \citep{2011PhRvL.107g1303Z}. This quantity,
\begin{equation}\label{d_def}
D_{N,f} \equiv \frac{d_{N,M_{\rm NB}/M_L \geq f}}{r_{\rm NB}},
\end{equation}
is defined as the distance to the $N$'th nearest neighbor whose mass exceeds $f$ times the halo under consideration divided by the virial radius of the neighboring halo. A large value of $D$ indicates that the halo lives in a low-density environment in the sense that it has no larger halos close by. It was shown in \cite{2011arXiv1103.0547H} that the quantity $D \equiv D_{1,1}$ represents the local density well and is almost uncorrelated with the halo mass. We have explicitly checked that this is also the case for our simulations.

We follow \cite{2011PhRvL.107g1303Z} and define a {\it high density environment} as $\log_{10}D < 1$ and a {\it low density environment} as  $\log_{10}D > 1$. Halos in a low- and high-density environments will be called {\it isolated halos} and {\it clustered halos} respectively. In order to study the variation of $\Delta_M$ with halo mass we will say that a halo with $M_L/(M_{\rm sun}/h) > 5\times 10^{12}$ is a {\it large halo}, while a halo with  $5\times 10^{11} < M_L/(M_{\rm sun}/h) <  2\times 10^{12}$ is a {\it small halo}. The lower limit comes from the fact that smaller halos in our simulations are not well resolved (less than $\sim$ 500 particles per halo) and will therefore not be used in this analysis. This choice, arbitrary as it might seem, is made so that we have approximately equal numbers of halos in each of the two categories.

\begin{figure*}
\centering
\includegraphics[width=2\columnwidth]{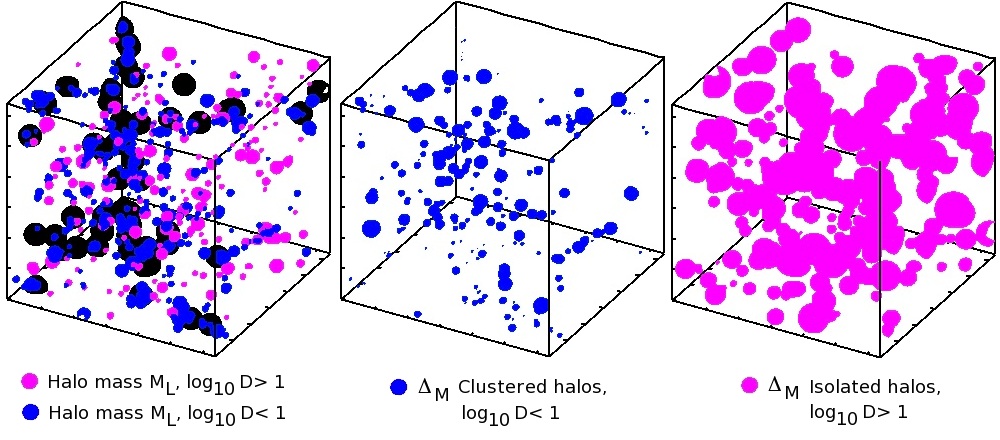}
\caption{The 3D distribution of the halos in the simulation box for a simulation with $z_{\rm SSB} = 0.5$ and $\beta = L = 1.0$. Left: The blue ($D<10$) and purple ($D>10$) spheres are 500 randomly selected halos in the mass range $11.5 < \log_{10} \frac{M_L}{M_{\rm sun}/h} < 12.5$. Because of the definition of the environment Eq.~(\ref{d_def}) the largest halos in the simulation will almost always be in a low-density environment and we therefore separate our the 50 most massive halos and show these separately in black with a fixed size for the spheres. Middle and right: The value of $\Delta_M$ for the clustered halos ($\log_{10} D < 1$) and the isolated halos ($\log_{10} D > 1$) respectively.}
\label{3dhalos}
\end{figure*}

In Fig.~(\ref{3dhalos}) we show the 3D halo lensing mass distribution together with the corresponding value of $\Delta_M$ in the simulation box for one of the simulations. Each sphere represents a halo; in the left panel the size of the spheres is proportional to the halo lensing mass $M_L$ and in the other two panels it is proportional to $\Delta_M$; the color indicates the environment of the halos as illustrated by the legend. This plot shows the environmental dependence clearly: in high density environments the value of $\Delta_M$ is generally smaller than in low density environments, and the definition of the environment used here is capable of capturing this behavior fairly well.

\section{Results}\label{sec4}

We use the N-body simulations of \cite{2011arXiv1108.3081D} which have been performed using a modified version \citep{baojiu_coupled_sfc} of the publicly available N-body code MLAPM \citep{mlapm}. The simulation suite consists of six simulations with different model parameters $\{\beta,L,z_{\rm SSB}\}$ shown in Table~(\ref{simdet}), and we have calculated $\Delta_M$ for the halos found in these simulations. The initial conditions for the symmetron models are the same for each simulation and allows for a direct comparison of the effects of the different parameters in the theory. The halos in the simulation have been found with the halo finder MHF \citep{Gill:2004km} using the definition $M_{\rm vir} = M(r_{340}) \equiv M_{340}$ for the halo mass, where $r_{340}$ is the radius of the halo where the local density is $\rho = 340 \overline{\rho}$.

\begin{figure*}
\centering
\includegraphics[width=2\columnwidth]{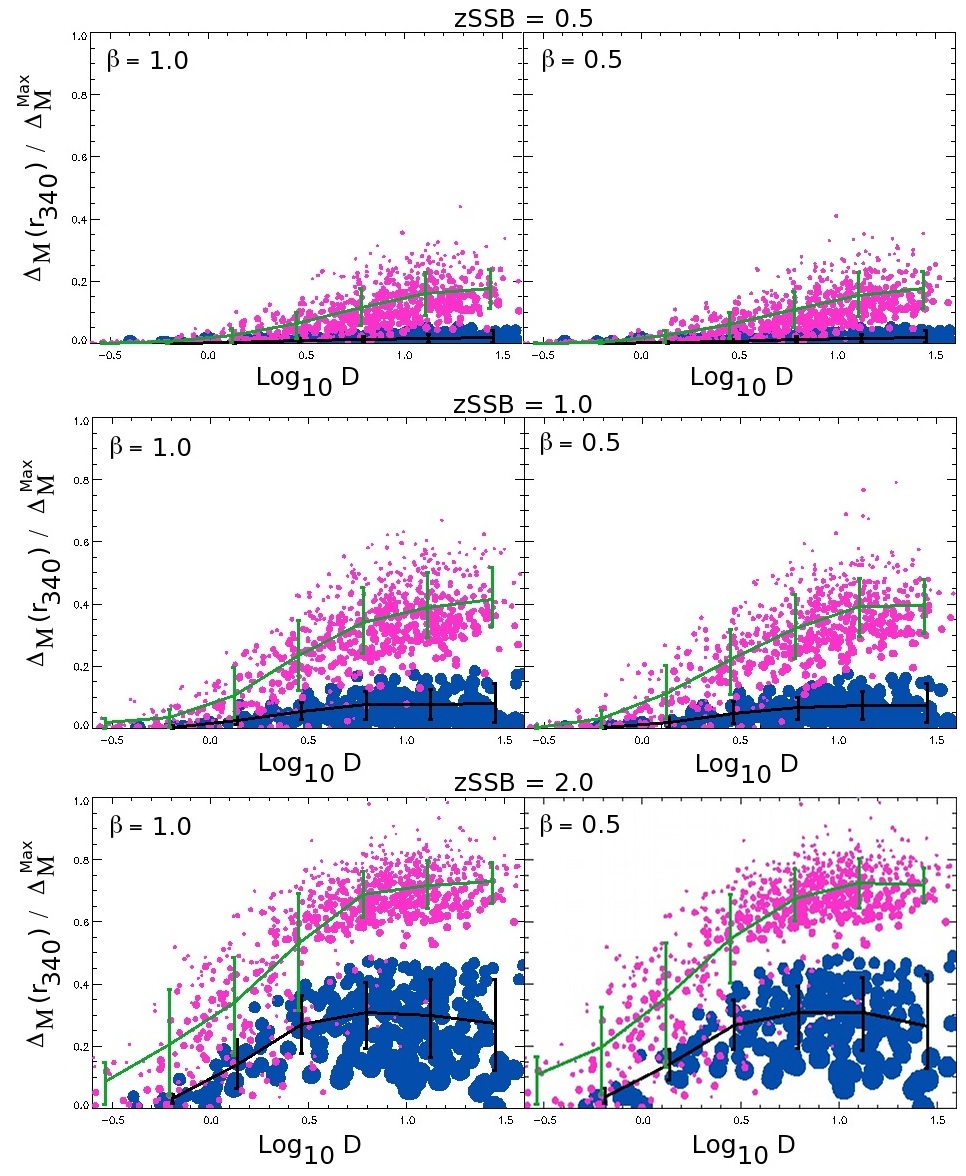}
\caption{$\Delta_M(r_{340})/\Delta_M^{\rm Max}$ for as function of the environment for large halos ($\frac{M_L}{M_{\rm sun}/h}>5\cdot 10^{12}$, blue) and small halos ($\frac{M_L}{M_{\rm sun}/h}<2\cdot 10^{12}$, purple) where $\Delta_M^{\rm Max} = 2\beta^2$ and where the size of the circles increases with the mass of the halos. The error bars are 1$\sigma$. We see a clear difference between the value of $\Delta_M$ between what we have defined as high-mass and low-mass halos.}
\label{deltaofD}
\end{figure*}

The condition for screening of an isolated halo (i.e. not taking the environment into account) can be found theoretically. For a spherical top-hat over density with radius $r_{340}$ and density $\rho = 340\overline{\rho}_{m0}$ we have
\begin{align}
\frac{\Delta_M(r_{340})}{\Delta_M^{\rm Max}} \simeq \left\{\begin{array}{cc}\alpha^{-1} & \alpha \gtrsim 1\\1 & \alpha \lesssim1\end{array}\right.
\end{align}
where $\alpha$ is given in Eq.~(\ref{alpha}). The condition for screening, $\alpha \gtrsim 1$, can be written as
\begin{align}\label{mass_screened}
\frac{M_{340}}{10^{12} M_{\rm sun}/h} &\gtrsim 0.6(1+z_{\rm SSB})^3L^2\left(\frac{r_{340}}{\text{Mpc}/h}\right),
\end{align}
where $M_{340}= \frac{4\pi}{3}\rho r_{340}^3$ is the halo mass. This condition is not accurate for real halos as the top-hat approximation is very crude \citep{2011arXiv1110.2177C}, but it nevertheless is able to capture the essence of the screening mechanism. In Table~(\ref{screened}) we show the ratio of halos more massive than $5\times 10^{11} M_{\rm sun}/h$ which are expected to be screened in the different simulations based on this simple approximation. The effectiveness of the symmetron screening mechanism increases with increasing halo mass $M_{340}$ and decreasing symmetry-breaking redshift $z_{\rm SSB}$. This is because a larger value of $z_{\rm SSB}$ means the symmetry is broken at higher matter densities and consequently a larger halo mass is required to restore the symmetry.
\\\\
In Fig.~(\ref{deltaofD}) we show $\Delta_M(r_{340})$ for our simulations as a function of the environment for both large (blue circles) and small (purple circles) halos. Firstly, we note that the predictions from simulations with different $\beta$ are very similar and the only real effect of changing $\beta$ is in changing the normalization factor $\Delta_M^{\rm Max} = 2\beta^2$. This can be understood from noting that changing $\beta$ only affects the geodesic equation Eq.~(\ref{geodet}) and not the Klein-Gordon equation\footnote{The Klein-Gordon equation Eq.~(\ref{eom_phi}) can be written as $\square\psi = \frac{1}{2\lambda_0^2}\left[\frac{\rho}{\overline{\rho}_{m0}(1+z_{\rm SSB})^3} - 1 + \psi^2\right]\psi $ where $\psi = \phi/\phi_0$. Using this variable the fifth-force can be written $F = \beta^2\left(\frac{M}{M_{\rm pl}}\right)^2 \psi \nabla\psi$. Thus the solution $\psi$ is independent of $\beta$ and its only the geodesic equation, through the fifth-force, which has a $\beta$ dependence.} Eq.~(\ref{eom_phi}). However, for simulations with larger $\beta$ we will on average have more massive halos because the fifth-force, and therefore the matter clustering, is stronger. This effect, which for our simulations is very small, can also be seen in Table~(\ref{screened}). Secondly, we note that the predictions of $\Delta_M(r_{340})$ for high-mass and low-mass halos in low-density environments are easy to separate at 1$\sigma$ for all our simulations. The small dispersion about the solid curves (which represent the average values in the two mass bands) seen in Fig.~(\ref{deltaofD}) is due to the difference in the halo masses within each mass band. 

To see this more closely, we have used the size of the circles (which represent halos) to denote their masses: bigger circles are more massive halos. We can see the clear trend that $\Delta_M$ decreases with increasing circle size (or halo mass), and this confirms that in a given environment the screening of a halo, or equivalently $\Delta_M$, depends very sensitively on the mass of that halo. Thirdly, for very high density environments $D\to 0$ we recover GR independent of the halo mass for all of our simulations, which is because the local value of $\phi$ in a very high density environment (which is often a part of or very close to a very massive halo) is small and so the fifth force is suppressed.

\begin{figure*}
\centering
\includegraphics[width=2\columnwidth]{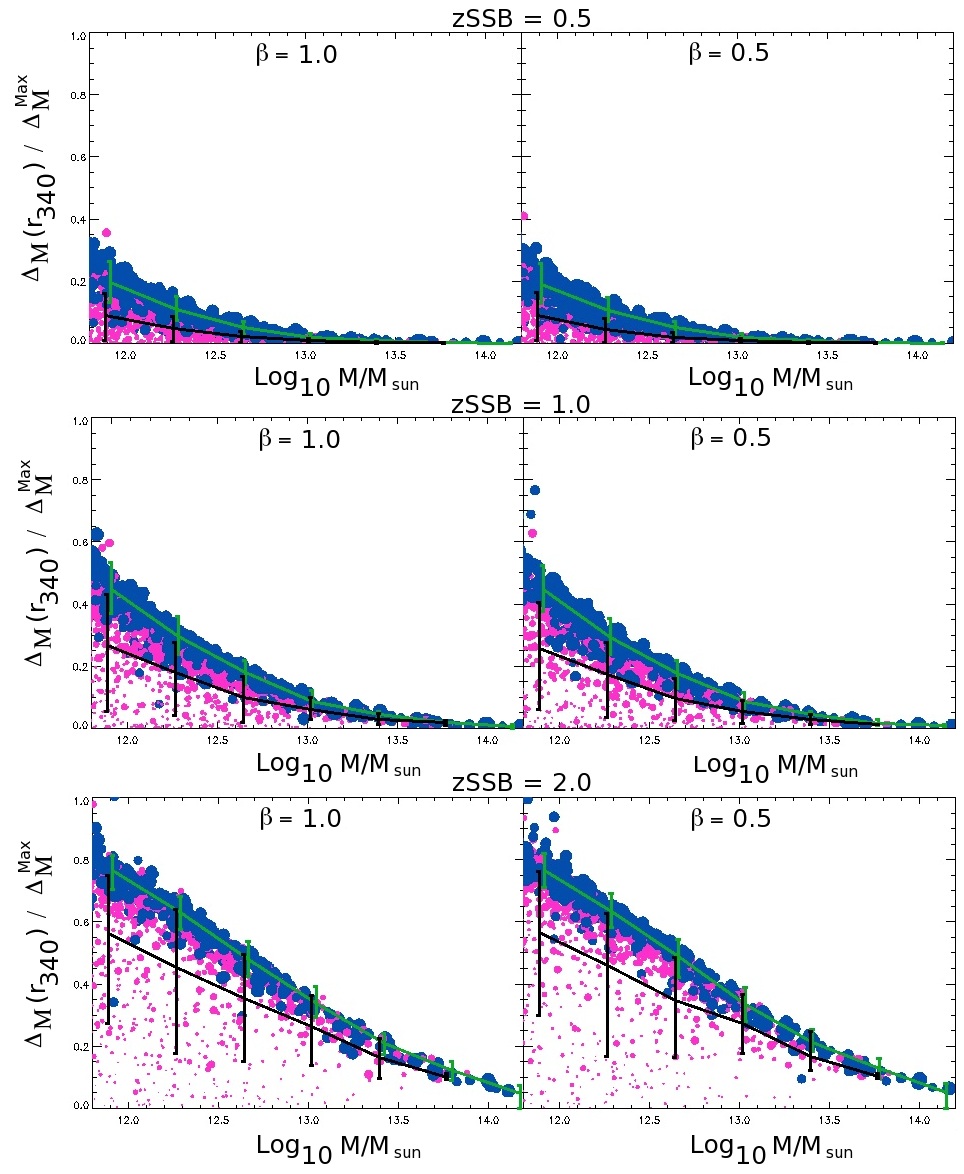}
\caption{$\Delta_M(r_{340})/\Delta_M^{\rm Max}$ as a function of the halo lensing mass $M_L$ for high density environments ($D<10$, purple) and low-density environments ($D>10$, blue) where $\Delta_M^{\rm Max} = 2\beta^2$ and where the size of the circles increases with $D$ (i.e. a smaller circle indicates a denser environment). The error bars are 1$\sigma$. For the high-mass halos we recover GR independent  of the environment as the effectiveness of the screening increases with mass.}
\label{deltaofM}
\end{figure*}

In Fig.~(\ref{deltaofM}) we show $\Delta_M(r_{340})$ for our simulations as a function of the halo mass in both high-density (purple circles) and low-density (blue circles) environments. This figure shows that GR is recovered for larger halos, independent of the environment, as expected from Eq.~(\ref{mass_screened}). For low-mass halos we see a significant dispersion of $\Delta_M$ from 0 to the maximum value obtained in low-density environments for the same mass-ranges. This is because low-mass halos cannot efficiently screen themselves and must rely on the environment to get the screening. The environment, defined by $D$, ranges from $D = 0$ up to $D=10$ for these halos, and the lower the value of $D$ the better screened the halo will be. To observe this point more clearly, in the figure we have also used the size of the circles to denote the value of $D$: the bigger circles are halos in environments with larger $D$ (or lower density) and vice versa. As expected, we see a clear trend that the small halos which are efficiently screened geneally reside in high-density environments, while those which are less screened lie in low-density environments.

Massive halos on the other hand can screen themselves efficiently and the environment only plays a small role in their total screening.

\begin{figure*}
\centering
\includegraphics[width=2.0\columnwidth]{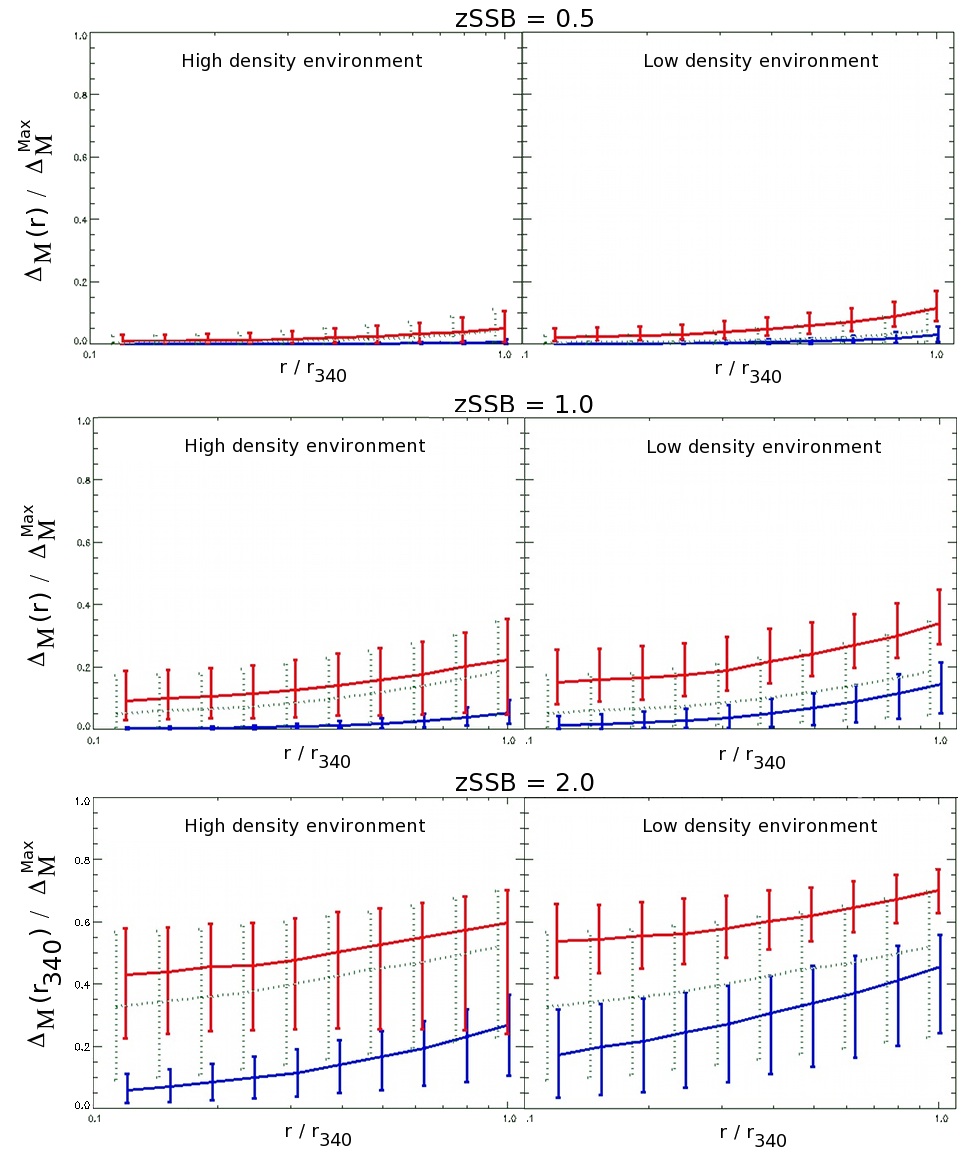}
\caption{$\Delta_M(r)/\Delta_M^{\rm Max}$ as a function of the rescaled halo radius $r/r_{340}$, where $\Delta_M^{\rm Max} = 2\beta^2$, for high ($D<10$) and low ($D>10$) density environments and small (red) and large (blue) halos. The error bars are
 1$\sigma$. For comparison we show the profile for all halos in the simulation (dashed green) in each plot and this curve has been displaced $-5\%$ in the $r$-direction to more clearly see the error bars.}
\label{deltaofr}
\end{figure*}

In Fig.~(\ref{deltaofr}) we show $\Delta_M(r)$ as a function of the distance $r$ from the halo centre, for small and large halos in high and low density environments respectively. Again we see a large difference between large halos in dense environments and small halos in low density environments. The $r$-dependence of $\Delta_M(r)$ is seen to be rather weak in high density environments since the value of the scalar field inside the halo is mainly determined by the environment, while in low-density environments the value of the scalar field mainly depends on the mass of the halo, which leads to a stronger $r$-dependence. Note also that in all the figures above the deviation from GR is stronger for higher symmetry-breaking redshift $z_{\rm SSB}$, as expected from Eq.~(\ref{mass_screened}), and for larger values of the coupling $\beta$, which implies a stronger fifth-force and therefore a stronger effect.

It should be emphasized that the environment dependence which are seen in the figures above will depend on the way the halos are binned, i.e. our definitions in Sec.~(\ref{env}).

\begin{table*}
\centering
\begin{tabular}{|l|c|c|c|c|c|c|}
\hline
 {\bf Simulation:} & {\bf A} &  {\bf B} &  {\bf C} &  {\bf D} &  {\bf E} &  {\bf F} \\
  \hline
   $z_{\rm SSB}$ 	& 0.5 & 0.5& 1.0& 1.0& 2.0& 2.0 \\
   \hline
  $\beta$ 			& 0.5 & 1.0& 0.5& 1.0& 0.5& 1.0 \\
  \hline
  $L$ 	 		& 1.0 & 1.0& 1.0& 1.0& 1.0& 1.0 \\
  \hline
\end{tabular}\begin{tabular}{|c|c|}
\hline
{\bf Parameter:} & {\bf Value:}\\
\hline
$\Omega_m$ & 0.267\\
\hline
$\Omega_{\Lambda}$ & 0.733\\
\hline
h & 0.719\\
\hline
$\sigma_8$ & 0.801 \\
\hline
$n_s$ & 0.963\\
\hline
$N_{\rm particles}$ & $256^3$\\
\hline
$B_0$ & $64\text{Mpc}/h$\\
\hline
\end{tabular}

\caption{The symmetron parameters for each of the N-body simulations A-F together with some relevant cosmological- and N-body parameters (same for all simulations). For a complete list of parameters see \cite{2011arXiv1108.3081D}.}
 \label{simdet}
\end{table*}

\begin{table*}
\centering
\begin{tabular}{|l|c|c|c|c|c|c|c|c|c|}
\hline
  {\bf Simulation:} & {\bf A} &  {\bf B} &  {\bf C} &  {\bf D} &  {\bf E} &  {\bf F}  \\
  \hline
{\bf Screened:} &99.82\% & 99.83\% & 64.38\% & 65.27 \% & 13.91 \% & 14.12 \% \\
  \hline
\end{tabular}
\caption{The percentage of halos more massive than $5\times 10^{11} M_{\rm sun}/h$ which are expected to be screened (to some degree) in the different simulations based on the approximation Eq.~(\ref{mass_screened}). There is a very small difference between the simulations where $\beta = 0.5$ (A,C,E) compared to $\beta = 1.0$ (B,D,F) even though the screening in only sensitive to $L$ and $z_{\rm SSB}$. This small difference comes from the fact that simulations with stronger $\beta$ will in general have more massive halos.}
 \label{screened}
\end{table*}

The results shown above are for halos at redshift $z=0$. Another signature in the symmetron scenario is the redshift dependence of $\Delta_M$. For halos at large redshifts, $z>z_{\rm SSB}$, we have $\Delta_M \approx 0$, independent of the environment and mass as the symmetry has not been broken at the background level and $\phi \sim 0$ almost everywhere in space, as demonstrated in Fig.~(7-10) in \cite{2011arXiv1108.3081D}. If this signature is found in observations, then dividing the observational samples in bins according to redshift one can probe the value of the symmetry breaking redshift $z_{\rm SSB}$.  The maximal strength of the deviation will again probe $\beta$, which can help to distinguish the symmetron from $f(R)$ gravity. If one for example finds $\Delta_M^{\rm Max} > \frac{1}{3}$ then $f(R)$ cannot account for the deviation.

\section{Comparison with analytical results}\label{sec5}
During the completion of this work a paper \cite{2011arXiv1110.2177C} came out with semi-analytical halo scale predictions for the symmetron. In their analysis they assumed a NFW profile and calculated the symmetron fifth-force for isolated halos. The quantity of interest is $\overline{g}_{\rm vir}$ (see \cite{2011arXiv1110.2177C,2010PhRvD..81j3002S} for details) which in our notation is given by
\begin{equation}
\overline{g}_{\rm vir} = 1 + \frac{\int r^3\rho(r) F_N\Delta_M(r)dr}{\int r^3\rho(r) F_N dr}
\end{equation}
where $F_N$ is the gravitational force. $\overline{g}_{\rm vir} $ is the average force to the average gravitational force over the halo. Since galaxies are spread around inside the halo, a measurement of the velocity dispersion of galaxies would therefore measure such an average. One of the cases shown in \cite{2011arXiv1110.2177C} can be compared to our simulation results, and as a consistency check we perform this comparison.

\begin{figure}
\centering
\includegraphics[width=\columnwidth]{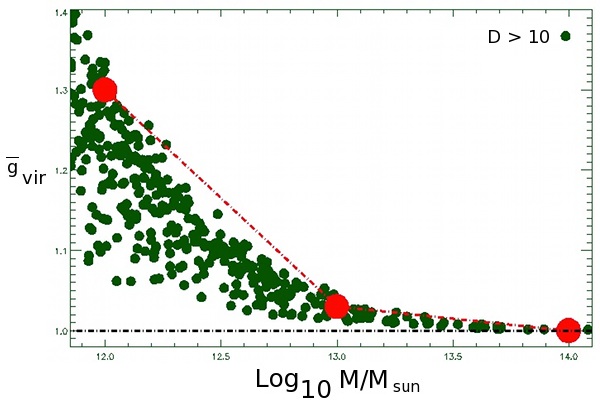}
\caption{$\overline{g}_{\rm vir}$ as function of halo mass for the isolated halos ($D>10$) in a simulation with $\{\beta=1.0, L=1.0, z_{\rm SSB} = 0.5\}$ compared with the semi-analytical results of (\cite{2011arXiv1110.2177C}, Fig.~3) (red). The dashed black line shows the GR prediction $\overline{g}_{\rm vir} = 1$. The semi-analytical results are for isolated halos which agrees very well to the maximum $\overline{g}_{\rm vir}$ in our simulations. Note that the symmetron parameters used in \cite{2011arXiv1110.2177C}: $\{\beta=1.0$, $L=1.2$, $z_{\rm SSB} = 0.54\}$ are slightly different compared to our simulation, and also for the definition for the virial mass $M = M_{300}$ was used apposed to our $M = M_{340}$.}
\label{comp}
\end{figure}

In Fig.~(\ref{comp}) we show $\overline{g}_{\rm vir}$ together with the predictions from \cite{2011arXiv1110.2177C}. The results from their analysis seem to be in good agreement with our numerical results. It would be interesting to see if their analysis can be extended by taking the environment into account to obtain the simulation results we have presented here. This would allow for an easier comparison with future observations as N-body simulations are in general very time consuming.

\section{Summary and conclusions}\label{sec6}

We have studied the environment dependence of the masses of dark matter halos in the symmetron modified gravity scenario. The potential governing the dynamics of the matter fields $(\Phi_- + \Phi_+)$ can differ significantly from the lensing potential $\Phi_+$ in this model, which leads to a clear difference between the mass of the halo as obtained from dynamical measurements and that obtained from gravitational lensing. Such an effect found in the symmetron model can be significantly stronger than in $f(R)$ gravity. This signature, which is unique to modified gravity, can in practice be measured by combining dynamical (e.g. velocity dispersion) and lensing mass measurements of clusters of galaxies or even single galaxies. We find that the environmental dependence is strongest for small halos as very large halos are sufficiently massive to be able to screen themselves. This implies that using dwarf galaxies \citep{2011arXiv1106.0065J} might prove the best way to probe this effect. 

This discovered feature of environmental dependence also allows us, in principle, to distinguish between different modified gravity scenarios such as $f(R)$, more general chameleons, DGP and the symmetron. In both DGP and $f(R)$ the maximum fraction of the fifth-force to the Newtonian force in halos are around $30\%$ while in chameleon/symmetron scenarios this fraction can be either smaller or larger, depending on the value of the coupling strength $\beta$. DGP differs from $f(R)$ and the symmetron in that there is basically no environmental dependence. There is also the possibility of measuring the redshift evolution of this effect by measuring clusters at high and low redshifts. As the symmetron force is negligible for $z > z_{\rm SSB}$ we will recover the GR predictions for all clusters, independent of mass and environment, at high redshifts.

Since different modified gravity theories can be highly degenerate with regards to both background cosmology and the growth rate of linear perturbations, it is crucial to identify new probes which can be used to separate them from each other. If one of these models are realized in nature then only a combination of many different probes will be able to pin down the correct theory. However, the first step would be to detect a deviation from GR and a detection of the effect considered in this paper will be a smoking gun for modified gravity. It will therefore be very interesting to look for this effect using data from upcoming large-scale structure surveys.

\section*{Acknowledgments}
The simulations used in this paper has been performed on TITAN, the
computing facilities at the University of Oslo in Norway. D.F.M. and H.A.W. thanks the Research Council
of Norway FRINAT grant 197251/V30. D.F.M. is also partially
supported by project PTDC/FIS/111725/2009 and CERN/FP/123618/2008. 
B.L. is supported by Queens' College, the Department of Applied Maths and
Theoretical Physics of University of Cambridge and the Royal Astronomical Society. H.A.W. thanks S. K. N${\ae}$ss for many useful discussions
\\\\
\newpage
\bibliography{env_dep_halo_symmetron}
\end{document}